\documentclass[a4paper]{article}

\usepackage[pages=all, color=black, position={current page.south}, placement=bottom, scale=1, opacity=1, vshift=5mm]{background}
\usepackage[margin=1in]{geometry} 

\usepackage{amsmath}
\usepackage{amsthm}
\usepackage{amssymb}

\usepackage[utf8]{inputenc}
\usepackage{hyperref}
\hypersetup{
	unicode,
	pdfauthor={Jonathan Bright, Florence E. Enock, Saba Esnaashari, John Francis, Youmna Hashem, Deborah Morgan},
	pdftitle={Generative AI is already widespread in the Public Sector},
	pdfsubject={Generative AI is already widespread in the Public Sector},
	pdfkeywords={Generative AI, Public Sector Productivity, Public Services},
	pdfproducer={LaTeX},
	pdfcreator={pdflatex}
}

\usepackage[sort&compress,numbers,square]{natbib}
\usepackage{graphicx, color}
\graphicspath{{fig/}}

\title{Generative AI is already widespread in the Public Sector}
\author{Jonathan Bright$^1$ \and Florence E. Enock$^1$ \and Saba Esnaashari$^1$ \and John Francis$^1$ \and Youmna Hashem$^1$ \and Deborah Morgan$^1$}

\date{
      $^1$Alan Turing Institute \\ \texttt{jbright@turing.ac.uk}\\[2ex]%
      \today
}

\begin{document}
	\maketitle
	
\begin{abstract}
Generative AI has the potential to transform how public services are delivered by enhancing productivity and reducing time spent on bureaucracy. Furthermore, unlike other types of artificial intelligence, it is a technology that has quickly become widely available for bottom-up adoption: essentially anyone can decide to make use of it in their day to day work. But to what extent is generative AI already in use in the public sector? Our survey of 938 public service professionals within the UK (covering education, health, social work and emergency services) seeks to answer this question. We find that use of generative AI systems is already widespread: 45\% of respondents were aware of generative AI usage within their area of work, while 22\% actively use a generative AI system. Public sector professionals were positive about both current use of the technology and its potential to enhance their efficiency and reduce bureaucratic workload in the future. For example, those working in the NHS thought that time spent on bureaucracy could drop from 50\% to 30\% if generative AI was properly exploited, an equivalent of one day per week (an enormous potential impact). Our survey also found a high amount of trust (61\%) around generative AI outputs, and a low fear of replacement (16\%). While respondents were optimistic overall, areas of concern included feeling like the UK is missing out on opportunities to use AI to improve public services (76\%), and only a minority of respondents (32\%) felt like there was clear guidance on generative AI usage in their workplaces. In other words, it is clear that generative AI is already transforming the public sector, but uptake is happening in a disorganised fashion without clear guidelines. The UK's public sector urgently needs to develop more systematic methods for taking advantage of the technology. 

		\noindent\textbf{Keywords:} Generative AI, public services, productivity
	\end{abstract}
    
\section{Introduction}

The economic and societal impacts of generative artificial intelligence (GenAI) systems have received significant attention \cite{dwivedi_artificial_2021}, but few studies have examined how such systems are used in the public sector, and how they are impacting operational public sector workers, or `street-level bureaucrats' \cite{lipsky_street-level_2010}. Given their accessibility and rapid proliferation, answers to these questions will be vital in helping us to safely and robustly actualise the potential of GenAI technologies in the public sector. We seek to tackle these questions and in this article.

GenAI systems have been made easily accessible, with features often available for free or included within corporate software packages. Interfaces such as ChatGPT or Microsoft Copilot represent tools that potentially any public sector worker can potentially use in their work however they wish, only needing an internet connection (though of course paid for versions of these technologies are emerging - however even then subscription prices are modest). This is a mode of AI deployment in the public sector which is completely different from the usual centralised corporate procurement process, where technologies are designed and deployed in a top down fashion. As such, wide use of these systems and integration into existing workflows may present a significant `bottom up' transition in the nature of public sector work at the micro-level, driven by the discretionary needs of street-level bureaucrats \cite{lipsky_street-level_2010}. Discretion has been found to have positive effects for service users, and is inherently valued by bureaucrats \cite{tummers_policy_2014}, meaning this mode of technology adoption might be more positive than that focussed on the top-down imposition of systems. Given the rapid speed of growth over the past year, the extent of GenAI use by public sector workers has understandably not yet been the subject of considerable research. We therefore seek to understand why and how public sector workers are using these systems, and more specifically, the impact of these systems upon their work and their own perceptions of productivity.

Our approach is based on a survey of public sector workers in the fields of healthcare, education, social care, and the emergency services within the UK. Respondents were asked about their use of different types of AI, their trust and understanding of these systems, as well as their concerns and optimism about this technology. We find considerable use of GenAI within the public sector, and high levels of optimism about the effect this technology might have on the future of public service delivery. Despite these findings, respondents also reported a lack of clarity regarding oversight, as well as low awareness of guidance on appropriate use. Importantly, despite there being wide variation in use cases for GenAI across professions, public service professionals were positive about the ability for AI to enhance productivity, and reported little worry that AI would eventually replace their current job.

\section{Background: AI and productivity in the public sector}

AI systems within public services seem to have the potential to increase public sector capacity in a range of areas \cite{margetts_rethink_2019, margetts_rethinking_2022}; and these systems are already in increasingly widespread use within the public sector. Such systems include, for example, the use of predictive analytic systems to support the allocation of resources in the area of health and social care, or the use of spatial analysis to create heat maps of residents' proximity to services in the area of planning and development \cite{bright_data_2019}. Many of these technologies have a productivity focus, with the aim being to free up the time of skilled public sector workers. However, despite a number of promising, large-scale public sector `digital transformation' initiatives and policies across government, recent UK Office for National Statistics (ONS) data has highlighted that total public service productivity only grew by an average of 0.2\%  per year between 1997 and 2019, with several service areas completely static or seeing diminishing growth \cite{office_for_national_statistics_public_2023}. In other words, up until today, the public sector has not taken much advantage of the major changes in digital technology that have been seen in the last two decades, or at least such changes have not been translated into enhanced productivity. 

The emergence of GenAI seems to create the opportunity to change this dynamic. The widespread deployment of freely available generative systems with readily accessible user interfaces places significant AI resources in the hands of operational - increasingly time pressed - public sector workers, who often spend a large amount of their day on bureaucracy; for instance Viechnicki and Eggers found that the average civil servant spends up to 30\% of their time on documenting information and other basic administrative tasks \cite{viechnicki_how_2017}. Clearly, automated support with such activities could support enhanced delivery of services and potentially re-orient civil servants work around tasks that require human intelligence (perhaps also resulting in higher job satisfaction \cite{berryhill_hello_2019}). This is supported by work in the private sector: research carried out by Brynjolfsson et al., has found that GenAI systems can positively impact productivity, particularly for novice and low-skilled workers \cite{brynjolfsson_generative_2023}. 

The potential for GenAI is increased, as we highlighted above, by the ease with which it can be adopted. A necessary dimension of the majority of existing and prior digital transformation projects is that they are conceptualised, procured, and implemented by centralised public sector bodies to address a defined problem. GenAI completely reverses this paradigm: its emergence and accessibility means that anyone with an internet connection can theoretically start to integrate it into their day to day working life. This bottom-up deployment creates huge potential for rapid (though likely uneven) adoption.

However, thus far we have only limited knowledge about the actual extent of this uptake in the public sector. Some statistics from early 2023 indicated that 8.2\% of employees at global companies had used ChatGPT \cite{noauthor_statistics_nodate}, while statistics from the UK indicated that more than a quarter of adults had used the tool \cite{milmo_more_2023}. But these statistics do not address the public sector. In terms of public sector more specifically, a recent report from the Department for Education showed that teachers were using it in a variety of capacities \cite{noauthor_ai_2023} (though without measuring the actual extent of usage), and recent survey work with the Canadian Federal Public Service found that 11.2\% had used generative systems for work purposes, with positive perceptions of the use of AI systems for data processing \cite{global_government_forum_attitudes_2023}. However, in large part due to the rapid deployment of such systems within the public realm, few studies have surveyed operational public sector workers across sectors regarding their use of systems to understand impacts upon perceptions of discretion, responsibility, and productivity. Such comparative work across sectors \cite{selten_just_2023} can support more nuanced and robust understandings of the impacts of GenAI as a professional tool. 

In addition to understanding levels of use of the technology, it is also worth understanding the extent to which people using the technology feel like they have clear guidance around it. One of the critical potential disadvantages of a `bottom-up' technology is that its adoption becomes somewhat chaotic, with different people using it in different ways, potentially undermining things such as the duty towards public sector equality. Of course, some guidance does exist. Earlier this year, the UK government published guidance to civil servants on the use of GenAI \cite{cabinet_office_guidance_2023} noting the risks around inputting sensitive and personal data alongside the risks of bias and misinformation from these systems. However, they also note that such systems can be helpful and assist with the work of civil servants, underscoring that they `should be inquisitive about new technologies, including generative AI tools.' \cite{cabinet_office_guidance_2023}. This government-wide guidance has been followed by sector specific guidance from individual departments (see e.g. \cite{noauthor_generative_nodate}). However questions remain about whether professionals are even aware this guidance exists, or the extent to which it is useful in their day-to-day engagements with the technology. Again, this is something we will seek to tackle in the present article.  

\section{Methods}

Data collection for this work took place online in November 2023 through a survey that was created and administered using Qualtrics\footnote{https://www.qualtrics.com}. Participants were recruited through Prolific\footnote{https://www.prolific.com}, and had to be based in the UK and over the age of 18 to take part. Furthermore, anyone taking part in the survey had to be working in one of five key public sector areas: the NHS, the emergency services, social work, schools or universities (it is worth noting of course that in the UK, universities are not entirely `public sector' in the sense that they are not owned by government, but they receive considerable amounts of funding from the public sector). Participants were paid at an equivalent rate of approximately £20 per hour for completing the survey. Participants were not allowed to complete the survey more than once. An attention check was included within the survey, with responses not recorded for those who failed the check.    

A total of 938 individuals completed the survey. Of the final 938 respondents, a majority (67\%) were female and most respondents reside within England (84\%). Respondents were fairly balanced across professions, with the largest number of responses received from NHS workers (24\%) and the fewest from emergency service workers (15\%). Time spent in the profession ranged from less than a year to 48 years, with a mean of about 11 years (SD=9.1), meaning that we capture the views of both junior and senior professionals. A full list of demographics can be seen in Table \ref{tab:demo}.

\begin{table}[!htbp] \centering 
  \caption{Demographics of Survey Respondents} 
  \label{tab:demo} 
\begin{tabular}{@{\extracolsep{5pt}} ccc} 
\\[-1.8ex]\hline 
\hline \\[-1.8ex] 
Category & n & frequency \\ 
\hline \\[-1.8ex] 
Use Any AI & 306 & 32.62\% \\ 
Use Generative AI & 209 & 22.28\% \\ 
Use Decision-Suppport AI & 92 & 9.81\% \\ 
Use Perceptive AI & 80 & 8.53\% \\ 
Female & 634 & 67.59\% \\ 
Age - Under 30          & 180 & 19.19\% \\ 
Age - 30-34             & 174 & 18.55\% \\ 
Age - 35-39             & 165 & 17.59\% \\ 
Age - 40-44             & 129 & 13.75\% \\ 
Age - 45-49             & 81 & 8.64\% \\ 
Age - 50-59             & 151 & 16.1\% \\ 
Age - 60 years and over & 58 & 6.18\% \\ 
Profession - NHS                & 228 & 24.31\% \\ 
Profession - Schools            & 220 & 23.45\% \\ 
Profession - Universities       & 187 & 19.94\% \\ 
Profession - Social Care        & 160 & 17.06\% \\ 
Profession - Emergency Services & 143 & 15.25\% \\ 
Country - England          & 788 & 84.01\% \\ 
Country - Scotland         & 85 & 9.06\% \\ 
Country - Northern Ireland & 22 & 2.35\% \\ 
Country - Wales            & 43 & 4.58\% \\ 
Years in Profession - 0-5   & 262 & 27.93\% \\ 
Years in Profession - 5-10  & 235 & 25.05\% \\ 
Years in Profession - 10-20 & 262 & 27.93\% \\ 
Years in Profession - 20+   & 179 & 19.08\% \\ 
\hline \\[-1.8ex] 
\end{tabular} 
\end{table} 

It is worth highlighting two key limitations of the survey. First, the sample is not likely to be fully representative of the population of public sector works in the UK. Although, as seen in Table \ref{tab:demo}, we have a good range of ages, genders, and different types of seniority, we would of course expect there to be bias in our sample towards those who make use of paid survey platforms. Furthermore, although we select people working in different professions, we do not know the professional role of people in our sample: for example someone working in a school might be in an HR or IT function, rather than working directly with children. 

Respondents were asked about three different types of AI systems which they may have encountered within their work, defined in Table \ref{tab:defs}. Asking about a variety of systems allows us to contrast uptake levels between traditionally top-down technologies and bottom-up ones. For each AI system, participants were asked about their general knowledge and use of these systems in their professions. If the participant indicated they used any of them, they were then asked about their trust and understanding of these systems. For the purpose of our research, we primarily focused on GenAI: tools such as ChatGPT which generate text or images based on prompts from a user. We therefore included some additional targeted questions around GenAI implementation in the workplace. All participants were then asked about their perceptions of AI use in the public sector, including areas of concern and optimism around what AI will portend for the future of public services. A full breakdown of the survey procedure can be found in \hyperlink{subsection.7.1}{the Appendix}.

\begin{table}[!htbp] \centering 
  \caption{AI Definitions} 
  \label{tab:defs} 
\begin{tabular}{p{0.2\linewidth}  p{0.35\linewidth}  p{0.35\linewidth} @{\extracolsep{5pt}}}
\\[-1.8ex]\hline 
\hline \\[-1.8ex] 
Type of AI & Definition & Example\\ 
\hline \\[-1.8ex] 
Generative AI & These are systems which can create text or images on your behalf, often on the basis of specific prompts. & ChatGPT, which can be used to help draft responses to emails or summarise text. \\ 
\hline \\[-1.8ex] 
Decision-Support AI & These are systems where machines are able to support decision-making by predicting, recommending, or prioritising outputs based on a set of inputs. & The use of resource allocation systems in hospitals that help triage patients arriving in the emergency room based on the symptoms they present. \\ 
\hline \\[-1.8ex] 
Perceptive AI & These are systems where machines are able to process sensory information such as visual and auditory inputs. & The computer vision technology used for facial recognition to assess whether an applicant's passport photo meets the criteria for a passport, or to authenticate an individual's identity. \\ 
\hline \\[-1.8ex] 
\end{tabular} 
\end{table} 

The survey results were exported to R, where we examined participants’ responses using descriptive statistics. We conducted chi-square tests of independence and one-way ANOVA tests to examine if there were significant differences in survey responses based on participants' characteristics.

\section{Results}

\begin{figure}[h!]
    \centering
    \includegraphics[width=1.0\linewidth]{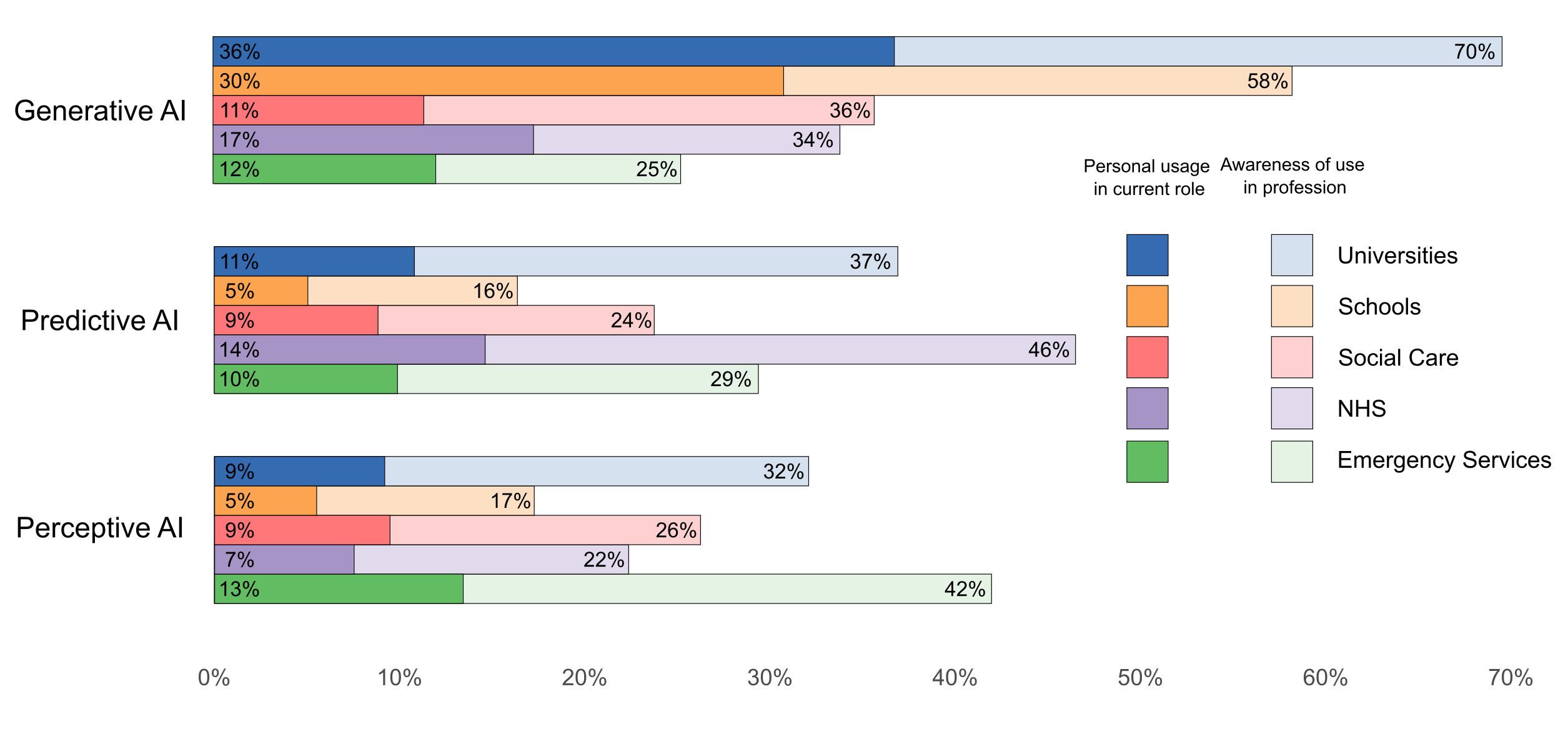}
    \caption{AI in the public sector}
    \label{fig:uptake}
\end{figure}

From Figure \ref{fig:uptake} we can see that GenAI use is already more widespread than predictive and perceptive AI applications in every profession apart form the emergency services, despite its nascent status as a consumer facing tool. Compared to other forms of AI with a longer history, such as facial recognition tools, GenAI has largely outstripped these other use cases in terms of uptake by the surveyed professionals. Over one fifth of respondents reported that they are currently using GenAI in their work, with respondents from Universities (36.4\%) and Schools (30.5\%) reporting the highest uptake. Respondents from the NHS (17\%), Emergency Services (12\%) and Social Care (11\%) reported lower uptake levels. The findings around awareness of other colleagues using the technology are also striking: 70\% of people in universities know at least one colleague using the technology, with high numbers for most of the other professions as well. 

In terms of demographic correlates of use, men are around 50\% more likely to use the technology than women, while increasing age lowers the likelihood of use (someone under 30 is about three times more likely to be using it than someone over 60). People who have been in their job a long time are, however, are just as likely to use GenAI as people who have only been there a short amount of time, once age differences are taken into account. Furthermore, GenAI is used frequently by those who have taken it up. Of those using the technology, almost 60\% are using it either on a daily or weekly basis in their work. 

Of the 209 respondents who reported that they use GenAI in their work, 132 (63\%) specified ChatGPT as the system they use, while the next highest specified system was Bard at 5\%. Respondents from Universities (26.7\%) and Schools (19.6\%) were most likely to have used ChatGPT in their work, while respondents from Emergency Services (5.6\%), NHS (8.8\%), and Social Care (6.9\%) were significantly less likely ($F=13.4, p<0.001$) to have mentioned use of ChatGPT. Notably, there was not a significant difference in uptake of ChatGPT by age or number of years in profession, although there was a significant difference by gender ($F=10.5, p<0.001$) , with male respondents (19.6\%) more likely to report using ChatGPT than female respondents (11.7\%).

\begin{figure}[h!]
    \centering
    \includegraphics[width=1.0\linewidth]{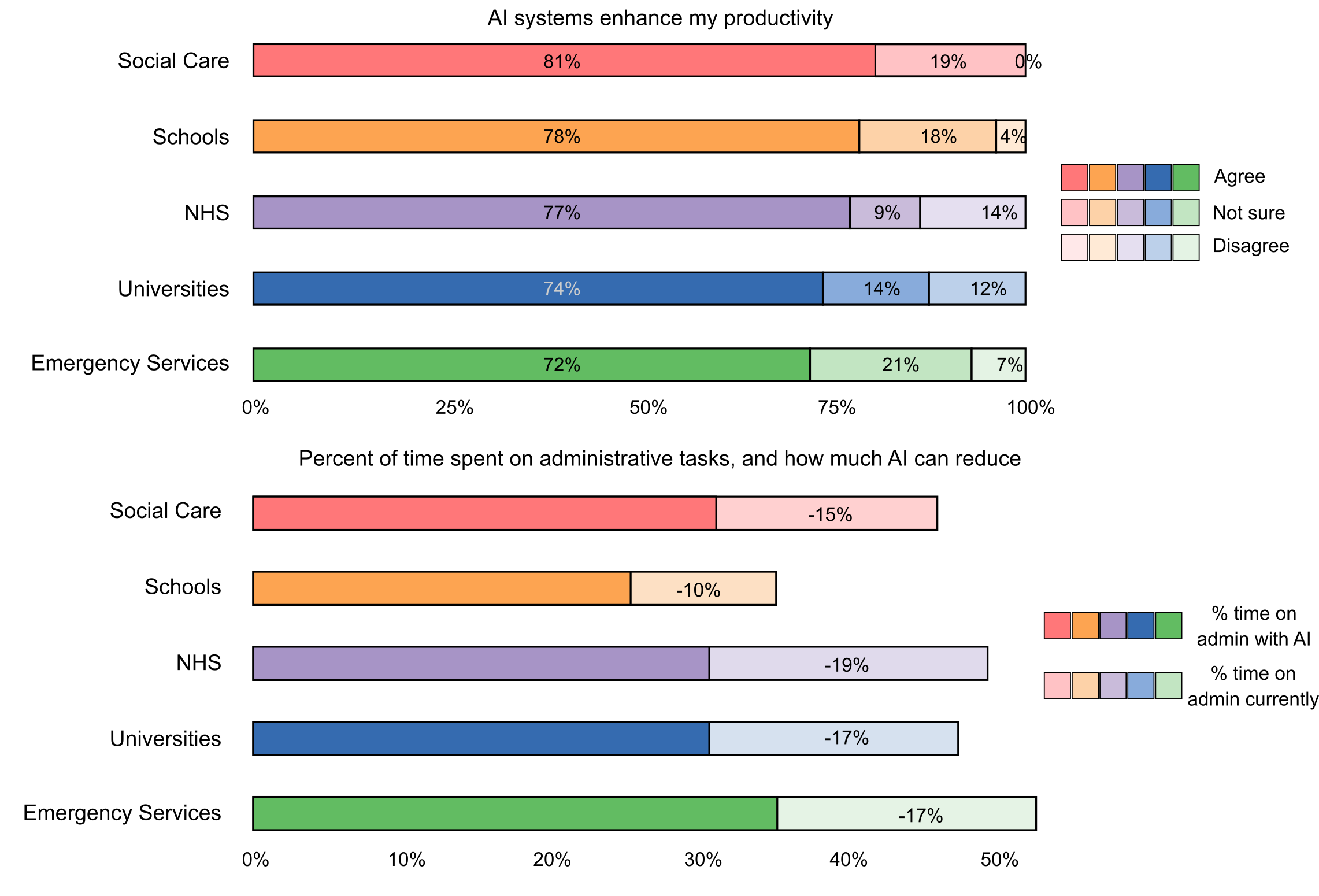}
    \caption{AI and productivity}
    \label{fig:bu_reduc}
\end{figure}

Respondents were positive about the use of AI for enhancing productivity. For example, more than 80\% of respondents in the social care profession said it enhanced their productivity, with other professions at similar levels. Figure \ref{fig:bu_reduc} shows that people are especially positive about the potential of AI to reduce bureaucracy in their work. The figure shows, on average, how much time people spend on bureaucracy, versus how much time they believe they would be spending if current AI technology was deployed effectively. All categories show a considerable decrease: for example, respondents working for the NHS indicated they spent almost 50\% of their time on bureaucracy, but thought AI use could reduce this to less than 30\% (the equivalent of saving an entire day of work every week). Respondents who reported that they use GenAI were 15\% more likely to agree that GenAI could reduce their time on bureaucracy. Additionally, the expected reduction in time spent on administrative tasks from GenAI tools was over 5\% higher on average (an additional two hours of time saved per week) from those who actively use these tools in their work. In other words, those using the tools can really see a benefit to them, and also can see potential to do more than is currently being done. 

When asked how they were using generative systems, respondent comments highlighted the utility of such systems to enhance productivity and creativity, and to reduce the time taken to complete tasks. Some ways respondents reported using GenAI were to draft emails, create examples and lesson materials, write up reports, aid in writing code, and generating synthetic data. Respondents specified a variety of ways in which GenAI was able to increase their productivity and was particularly useful to managing their workloads. One such response noted that, `I use ChatGPT, when I am overwhelmed with work', while another respondent mentioned that `as an IT technician, I utilize ChatGPT to streamline customer support, providing quick responses to common queries'. In terms of more creative tasks, one respondent noted that ChatGPT was useful `to generate ideas for marketing content, blog posts etc', while a teacher reported that ChatGPT and Bard were useful to `generate ideas for improving my teaching and supporting students'. The variety of responses received demonstrate the complexity and breadth of use of GenAI systems as an individual tool across professions in the public sector. 

\begin{figure}[h!]
    \centering
    \includegraphics[width=1.0\linewidth]{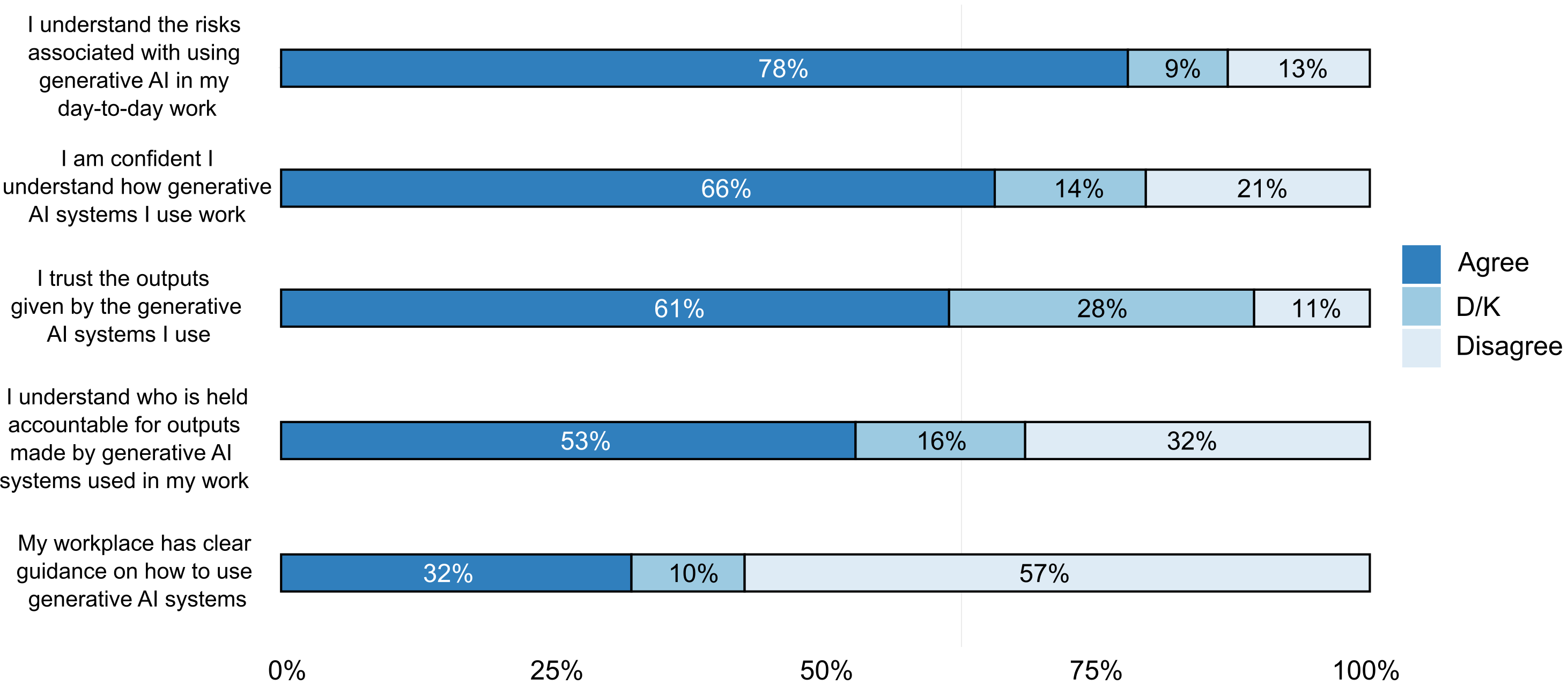}
    \caption{Understanding of Generative AI}
    \label{fig:under_AI}
\end{figure}

Respondents that use GenAI were overall quite trusting of AI technologies in the workplace (see Figure \ref{fig:under_AI}). For all professions, more than 60\% of GenAI users said that they understood how these AI systems worked. A majority of these respondents (over 60\% in all cases, except for Universities) also trusted the output of GenAI technology. Additionally, a large majority of respondents agreed that they understand the risks associated with using GenAI in their workplaces. Despite their use of GenAI, fewer respondents (53\%) said that they understand who is held accountable for the outputs produced by systems like ChatGPT. Furthermore, most professionals felt that the guidance around the use of GenAI in their workplace was not clear.

Concern about respondents' jobs being replaced by AI was not widespread. Only about 16\% of respondents were worried that AI would eventually replace their current job. This number was highest among people working at a university (25\%), although interestingly there was no significant difference in responses between people that use or don't use GenAI in their work. Overall, public service professionals responded positively about the potential benefits of AI use, with over 75\% of respondents optimistic about how AI will improve public services in the future. Respondents that use GenAI were significantly more likely to be optimistic ($t=4.42, p<.001$) than other respondents, with 86\% reporting optimism, versus only 72\% of non GenAI users. Despite this optimism, most respondents (76\%) agreed that the UK is missing out on opportunities to use AI to improve public services. This number was significantly higher ($t=3.65, p<0.001$) for users of GenAI (85\%) than for those who do not use AI in their work (73\%).

\section{Discussion}

Through a preliminary survey of public professionals in the UK, we find that use of GenAI already prevalent within the public sector. Respondents had high levels of optimism about the potential for GenAI to reduce the amount of time they spend on bureaucracy, and they believe this technology can improve public service delivery in the future. Despite these anticipated improvements to productivity, public sector workers are not particularly worried that this technology will eventually replace their jobs. Respondents also noted that despite their usage of GenAI tools, there is a general lack of guidance on appropriate use provided by their employers.

The results presented in this survey suggest that uptake of GenAI in the public sector is proceeding relatively rapidly. Despite only being available as a consumer facing product for around a year, GenAI use is more widespread than other types of AI that have been around comparatively longer. However, while uptake of GenAI is relatively high given the recent proliferation of this technology, our results also demonstrate that not everyone is using the technology. Indeed, in some public service professions, directly reported uptake remains only around 10\%. Considering the positive results reported by those who are using the technology, this seems like an opportunity for the public sector to push further, and ensure that it is rolled out more widely. 

Of course, there may be barriers to more widespread uptake. First, it is likely that not all jobs support easy implementation of GenAI tools, as suggested by Halal et al. \cite{william_halal_forecasts_2016}. One area where further research is needed is understanding the proportion of public sector workers who could benefit from using these tools. Furthermore, there may also be reluctance to implement GenAI in some areas of work. A recent survey of public attitudes to AI in the UK found that the public holds nuanced views on what it regards as proper use cases of AI\cite{ada_lovelace_institute_how_2023}. The same survey found that while the general public views efficiency and improved accessibility as a main advantage of AI, there is worry about AI being used to replace professional judgements, such as within hiring \cite{ada_lovelace_institute_how_2023}. While uptake of GenAI is high, it could also be that public attitudes towards this burgeoning technology are causing some public professionals to be a bit more cautious about integrating GenAI tools into their daily practice. Furthermore, we find that responsibility for the outputs of GenAI is currently unclear to public professionals, which echoes conclusions drawn from Brown, who charts the complications created by GenAI in AI supply chains, and discusses potential frameworks for addressing this issue \cite{brown_expert_2023}. Notably, issues of responsibility in workflows are not specific to AI, as other work has noted the difficulties in attributing responsibility or legal liability to other forms of complex supply chains \cite{cobbe_understanding_2023}. Lack of clear lines of responsibility, and a perception of a lack of clear guidance, could be a further factor blocking uptake. 

Despite these barriers, the levels of optimism we found around the technology was striking. One reason for this could be precisely the bottom-up way in which it has been adopted. GenAI provides enhanced personal agency in how it is used, with the ability to tailor and personalise use cases to meet distinct needs, rather than forcing users to make use of a system in a specific way. Our findings around time spent on administrative tasks suggest that there is widespread hope from professionals that AI will reduce the amount of time spent on this type of work. Despite this, there is little worry that professionals' jobs will be on the line. The future imagined by these respondent beliefs is one where AI will not eliminate human activity in the public sector workforce, but instead free up workers to spend more of their time on non-routine tasks. The high levels of reported optimism around the technology's potential to alleviate bureaucratic workloads echo similar sentiments in the United States, where research found that the majority of public sector professionals were optimistic about the role that AI can play in `improving bureaucratic efficiency' \cite{alva_impact_2021}. Alternatively, if, as the NHS workers in our sample predict, a full day of work currently taken up by administration can be eliminated, current productivity levels could be maintained while making a powerful case for moving to a four-day work week, the benefits and risks of which have long been debated \cite{campbell_four-day_2023}.

The widespread use of generative applications such as ChatGPT suggests that public sector workers may have already started becoming reliant on these technologies to automate common administrative tasks. This does raise the question of the business model behind these technologies, and who will provide them in the future. There remains a potential danger of a few key companies with the resources to create powerful GenAI tools having a stranglehold over the productivity of governments, and the workforce more broadly. However, GenAI technology is also becoming more widely distributed and accessible to create, with projects such as BLOOM \cite{scao_bloom_2023} attempting to democratise access to this technology. One key question for government moving forward will be around whether it wishes to invest in creating its own language models, supported by open source technology, or whether it will focus on procurement from technology companies. Resolving this question will be crucial to the future of generative AI uptake in the public sector.

\bibliography{genai_article}

\appendix
\section{Appendix}
\subsection{Survey Procedure}
For each participant, we first collected standard information about age and gender, along with public service profession, time spent in profession and time spent in current role. Participants next read a short definition of perceptive AI systems and were then asked if they were aware of such systems in use in their area of work (Yes/No/I don’t know). If participants responded `Yes' to awareness, they were then asked whether they make use of any of these perceptive AI systems (Yes/No/I don’t know). If they responded `Yes' to using such systems, they were asked to provide the name(s) or a brief description of the system(s) they use in a few words using a free text box, and they were then asked how often they use such systems (Daily/Weekly/Every month/Only once or twice before). In the next block, participants read a short definition of decision-support AI systems and then responded to the same questions about awareness, use and frequency of use as described for perceptive AI. Following these questions, participants read a short definition of generative AI systems and again responded to the same questions about awareness, use and frequency of use. 

If participants indicated that they use generative AI systems, they were then asked a follow up set of questions in which they were asked to indicate agreement with the following statements: `My workplace has clear guidance on how to use generative AI systems'; `I trust the outputs given by the generative AI systems I use'; and `Using generative AI has changed the scope of my day-to-day work' (Strongly agree/Agree/Neither agree nor disagree/Disagree/Strongly disagree/I don’t know). 

Next, if participants had indicated that they use more than one type of the three AI systems asked about, they were then asked to indicate which one they use most frequently (Perceptive/Decision support/Generative). 

If participants indicated that they had used at least one of the three system types before, they were presented with another set of follow up questions about their use and understanding of AI used in their current role. If they had indicated they use more than one system, they were asked to consider the one they use most often. Participants were asked to indicate agreement with the following four statements: `I am confident I understand how the AI systems I use work'; `I feel AI systems enhance my productivity'; `I understand the risks associated with using this AI system in my day-to-day work'; `I understand who is held accountable for outputs made by AI systems used in my work' (Strongly agree/Agree/Neither agree nor disagree/Disagree/Strongly disagree/I don't know). 

All participants were then asked to indicate approximately what percentage of their working week they typically spend on bureaucracy and administration, on a sliding scale from 0 to 100, and whether they think AI has the potential to decrease the amount of time spent on such tasks (Yes/No/Not sure). If they answered Yes, they were asked to indicate what percentage of the time spent on administrative tasks they think AI can help reduce on a sliding scale from 0 to 100. 

All participants were then asked a final set of questions relating to their general perceptions of AI use in the public sector. Participants were asked to indicate level of agreement with the following three statements: `I am concerned that AI will eventually replace my current job'; `I think we are missing out on the opportunities to use AI to improve public services'; `I am optimistic about how AI will improve public services in the future' (Strongly agree/Agree/Neither agree nor disagree/Disagree/Strongly disagree/I don’t know). 

Participants then responded to a simple attention check at the end of the survey and were then thanked and redirected back to Prolific for payment.

\subsection{Funding}
This work was supported by Towards Turing 2.0 under the EPSRC Grant EP/W037211/1 and The Alan Turing Institute. 

\end{document}